\pdfoutput=1
\documentclass[10pt,a4paper,twocolumn,superscriptaddress]{revtex4-1}
\usepackage[utf8]{inputenc}
\usepackage{amsmath}
\usepackage{amsfonts}
\usepackage{amssymb}
\usepackage{graphicx}
\usepackage{times}
\usepackage{microtype}

\usepackage[small,compact]{titlesec}
\usepackage[top=0.9in, bottom=1in, left=0.5in, right=0.5in]{geometry}
\begin{document}
	\title{Uncertainties in the Static Dielectric Constants computed from Molecular Dynamics Simulations}

    \author{Hernán R. Sánchez}
	\email{hernan.sanchez@quimica.unlp.edu.ar}
	\affiliation{Instituto de F\'{\i}sica de L\'{\i}quidos y Sistemas Biol\'{o}gicos. CONICET  La Plata - UNLP, Argentina.}
	
	\onecolumngrid

	\begin{abstract}
	In this work, the uncertainties in the dielectric constants of polar liquids, computed using molecular dynamics (MD) simulations, are compared for two different calculations schemes. Expressions of the uncertainty are derived for the external field method, and compared with those of the fluctuation method.  Significant differences on their the system size dependence were found.	In addition, alternative calculation procedures are  proposed. The individual contributions of different parts of the system to the electric susceptibility, and their corresponding uncertainties, are also studied. Additionally, the effects of the sampling frequency on the uncertainties are analyzed. MD simulations of pure liquid water (SPC/E) at 298.15 K and 1 bar were performed in order to corroborate the obtained results. In such conditions, the best estimate of the dielectric constant obtained in this study is 70.46~$\pm$~0.31.

\end{abstract}
	
	\maketitle
	
	\section{Introduction}
    The importance of the dielectric properties has been recognized for decades\cite{von1955dielectric}. Their measurement is useful for many application in many fields, for example, in the characterization of heterogeneous systems\cite{asami2002characterization}. The influence of the relative (static) dielectric permittivity in the interaction between particles is significant for condensed phases. Due to this, in order to get accurate values for many properties, it must be accurately represented when performing computer simulations of physical systems.
    
    For isotropic and homogeneous liquid systems, the dielectric constant can be computed from the total dipole moment. It in turn, can be obtained, among many alternatives, from molecular dynamics (MD) simulations\cite{alder1959studies,rapaport2004art}. The expressions that link the dipole moment with the dielectric constant depend on the boundary conditions. Those used in this work are valid for the tin-foil\cite{levrel2008boundary} and the Ewald\cite{de1986computer} boundary conditions, which are very common choices.
    
    There are many routes of calculation of the static dielectric constant ($\varepsilon_r$) of these kinds of systems \cite{heinz2001comparison}. Two of them are the most used in the literature. In one, the dielectric constant is related to the fluctuations in time of the total dipole moment of the system ($\boldsymbol{M}$), in absence of external fields, via the following equation\cite{neumann1983dipole,kolafa2014static}

    \begin{equation}\label{EQ FR}
	\varepsilon_r = 1 + \frac{\operatorname{Var}[\boldsymbol{M}]}{3\varepsilon_0 Vk_B T}
	\end{equation}
	
	\noindent where $\varepsilon_0$ represents the vacuum permittivity, $k_B$ the Boltzmann's constant, $V$ and $T$ system equilibrium volume and temperature, respectively. $\operatorname{Var}[\cdot]$ represent the variance operator. This calculation scheme will be referred here as the Fluctuation Route (FR).
	
	The other calculation route requires the application of an external  time-independent electric field ($\mathbf{E}$)\cite{riniker2011calculation}. If the direction of the field is given by an unitary vector $\check{\mathbf u}_E$, the dielectric constant can be obtained from\cite{neumann1983dipole,kolafa2014static}
			
	\begin{equation}\label{EQ EFR}
\varepsilon_r =  1 + \frac{\operatorname{E}[\boldsymbol{M}\cdot \hat{\boldsymbol{u_E}}]}{\varepsilon_0 V \boldsymbol{E}\cdot \hat{\boldsymbol{u_E}}}= 1 + \frac{\operatorname{E}[M]}{\varepsilon_0 V E}
\end{equation}

    \noindent where $\operatorname{E}[\cdot]$ represents the expected value operator, and $M$ and ${E}$ the euclidean norm of $\boldsymbol{M}$ and ${E}$, respectively.  Hereafter, this scheme of calculation will be referred as the External Field Route (EFR).
    
    Molecular dynamics can be used for sampling the total dipole moment. As the number of simulation steps is finite, $\operatorname{Var}[\boldsymbol{M}]$ and $\operatorname{E}[M]$ can only be estimated. Unfortunately, obtaining converged values of the dielectric constant require uncommonly large simulation times. Because of this, it is much more difficult to calculate $\varepsilon_r$ than many commonly computed properties\cite{caleman2011force}.    
    
    From a practical perspective, the dielectric constant cannot be obtained exactly, and commonly one of the following estimators is used,
    
    \begin{align}\label{eq: FR estimator}
    	\varepsilon_r &= 1 + \frac{\bar{\boldsymbol{M}^2}-\bar{\boldsymbol{M}}^2}{3\varepsilon_0 Vk_B T}\\\label{eq: EFR estimator}
    	\varepsilon_r &= 1 + \frac{\bar{M}}{\varepsilon_0 VE}
    \end{align}
		
   \noindent where the bar above a symbol denotes its sample arithmetic mean. Many times $\bar{\boldsymbol{M}}^2$ is not included in Eq. \ref{eq: FR estimator}\cite{neumann1983dipole}.
   
   In a previous work by the author an others, the analytical expressions for the variance of the dielectric constant computed through the FR (Eq. \ref{eq: FR estimator}) were found\cite{SANCHEZ2019546}. In this work, analytical expressions for the variance corresponding to the estimator of Eq. \ref{eq: EFR estimator} (EFR) are derived, and both calculations routes are compared in terms of their corresponding  uncertainties. These calculation schemes were compared before in terms of accuracy though numerical experiments, for example in Ref. \cite{kolafa2014static}. However, the usage of analytical expressions provides clear advantages.
   
   In the present study is also shown, that under usual circumstances, both calculation schemes can be combined for providing greater precision. This can be done by performing a MD simulation with an external electric field, computing the dielectric constant through the EFR, and using the information related to the remaining perpendicular directions in conjunction with the FR.
   
   Afterwards, a new route of calculation is presented. It requires a simulation without an external field, and uses the equation for the dielectric constant corresponding to de EFR (Eq. \ref{eq: EFR estimator}). It can be considered an intermediate step between the EFR and FR, as it arises from the derivation of the FR from the EFR. The proposed scheme provides no added value in terms of accuracy, as its final values coincide with those from the FR. However, the use of the former is advantageous for the study of the contribution to the electric susceptibility of different parts of the system (partial susceptibilities) and their corresponding uncertainties.
   
   The frequency at which the system configurations are stored is significant for the uncertainty in the computed dielectric constant. Although this has been studied before by means of numerical tests\cite{gereben2011accurate}, it remains unclear in the specific literature.  Because of this, the issue is addressed in this work.
	
	\section{Theory}
	The largest part of this work is focused in the calculation of uncertainties. The total dipole moment of the system obtained with MD is auto-correlated, i.e, the latter correlates with a delayed copy of itself\cite{gordon1968correlation}. This has huge impact in the estimations of the uncertainties of the mean and sample variance, as is discussed in the sections \ref{sec: registration} and \ref{sec: Uncert FR}.
	
	The auto-correlation function ($\phi(t)$) is a measure of the degree of correlation, of the signal with itself, as a function of the time delay. Its definition may depends on the context, in this work the following definition is used\cite{brockwell1991time}
	
	\begin{equation}\label{eq: ACF definition}
	\phi(t) = \frac{\operatorname{E}[(\mathbf M(t^\prime)-\operatorname{E}[\mathbf M(t)])(\mathbf M(t-t^\prime))-\operatorname{E}[\mathbf M(t)])  }{\operatorname{Var}[\mathbf M(t)]  }
	\end{equation}

    The definition above reflects the strength of the statistical dependence at different times, as it is normalized. The auto-correlation function is significant for this study, because the former appears in the derived expressions for the uncertainty of the dielectric constant. Many estimators of the auto-correlation function exist. Their differences are not significant for this work, in which the most popular discrete estimator was used, and is defined by

 \begin{equation}\label{eq: autocorr estimator}
 r(k\Delta t) = \frac{\sum_{i=1}^{n-k} (\mathbf M(i\Delta t)-\bar{\mathbf M})(\mathbf M((i+k)\Delta t)-\bar{ \mathbf M}) }{\sum_{i=1}^{n} (\mathbf M(i\Delta t)-\bar{ \mathbf M})^2}
 \end{equation}

    The dielectric permittivity can also be studied and modeled as a function of the frequency of the external field applied\cite{hilfer2002h}. Those representations in frequency domain have their counter-part in the time domain\cite{bottcher1978theory}. The expression in time domain describe the dielectric relaxation process. It can be shown that, taking aside constants, the time domain expression coincide with the auto-correlation of the total dipole moment of the system. This is significant for this work because it allows to approximate the auto-correlation function with well known expressions of the dielectric relaxation in time domain. Among them, the Debye's relaxation model\cite{hilfer2002h} will be used due to its simplicity. Its relaxation function which describes the decay of polarization is\cite{garrappa2016models}

\begin{equation}\label{ec: fx debye}
f(t) = e^{-t/\tau}
\end{equation}

\noindent where $\tau$ represents the relaxation time.

	\subsection{Simulation length and sampling frequency\label{sec: registration}}
	As mentioned above, computing the dielectric constant requires very large simulation times. This imply that huge file sizes will be obtained if the simulation coordinates are written to disk at every simulation step, or at least very frequently. In such cases, the subsequent processing of the full information may be computationally expensive. Fortunately, it is not necessary as the additional information provided for correlated steps is often less useful than the corresponding to uncorrelated steps. When observations are uncorrelated, the variance of the mean of a random variable $M$ satisfies $\operatorname{Var}[\bar M]= \operatorname{Var}[M]/n$, where $n$ represents the number of observations. It is not true for autocorrelated variables, however, a similar equation is obtained by defining the effective number of observations as\cite{bayley1946effective}

	\begin{equation} \label{eq: def neff}
	n_\text{eff}:= \operatorname{Var}[M]/\operatorname{Var}[\bar M]
	\end{equation}
	
	\noindent which is given by

	\begin{equation}\label{eq: equiv neff}
	n_\text{eff} = \frac{n}{1+2\sum_{i=1}^{n-1}(1-i/n)\phi(i\,\Delta t ) }
	\end{equation}
	
	For the cases taken into account in this paper, a good approximation is to consider the Debye's model.  This implies that $\phi(t)= e^{-\frac{t}{\tau}}$. Then, if $\tau << \mathcal{T} := n \Delta t$, we can approximate $1-i/n\approx1$. If both substitution are made in Eq. \ref{eq: equiv neff}, and $\Delta t$ is very small
    
    \begin{align}\label{eq: neff and t}
    n_\text{eff} &=\lim\limits_{\Delta t \rightarrow 0}  \frac{n\Delta t}{\Delta t+2\sum_{i=1}^{n-1}e^{-\frac{i\Delta t}{\tau}}\Delta t}\\\nonumber
    &=\frac{\mathcal{T}}{2\int_{0}^{\mathcal{T}}e^{-\frac{t}{\tau}}d t}=\frac{\mathcal{T}}{2\tau}
    \end{align}
	
	An alternative and elegant derivation of the above equation was previously found, for auto-regressive processes\cite{brockwell1991time}, in the reference \cite{zikeba2010effective}. In that work, it was also shown that for the process studied here
	
	\begin{equation}\label{eq: reg freq}
	n_\text{eff}= n \operatorname{tanh}\frac{\mathcal{T}}{2n\tau}
	\end{equation}
	
	This equation should be taken into account when deciding the sampling frequency. It is also significant in that it may serve to estimate the applicability of Eq. \ref{eq: neff and t}. The proposed derivation can be adapted to other relaxation processes which are not well represented by the Debye's model. The Eq. \ref{eq: equiv neff} cannot be used in practice because $\phi(i\,\Delta t )$ is not known. However, it can be estimated from a small sample and the quotient $n_\text{eff}/n$ may help in choosing a sampling frequency in a more general cases.

	\subsection{Uncertainty in the dielectric constants through the external field route\label{sec: uncert EFR}}

    Applying the variance operator to Eq. \ref{eq: EFR estimator}, and combining the result with Eq. \ref{eq: def neff}, the following expression is obtained
	
	\begin{equation}\label{eq: uncert EFR}
	\operatorname{Var}[\varepsilon_r]= \frac{1}{(\varepsilon_0VE)^2} \frac{	\operatorname{Var}[M]}{n_\text{eff}}
	\end{equation}
	
	The Eq. \ref{eq: uncert EFR} allows to evaluate the uncertainty in the dielectric constants computed though the EFR. The standard deviation turns to be inversely proportional to the external electric field, as stated previously in reference \cite{kolafa2014static}. When the relaxation times are known, the Eq. \ref{eq: uncert EFR} can be useful for the estimation of the simulation length required for a given target uncertainty, as it can be rewritten as
	
	\begin{equation}\label{eq: uncert EFR times}
	\operatorname{Var}[\varepsilon_r]= \frac{1}{(\varepsilon_0VE)^2} \frac{2\tau_E	\operatorname{Var}[M]}{\mathcal{T}}
	\end{equation}
	 
	 \noindent where the subscript $E$ was included in $\tau_E$ for stressing the fact that, because of the external electric field, this is not the exact relaxation time of the substance. However, it is reasonably close to the true relaxation time for the present analysis. Notice that employing the Eq. \ref{eq: uncert EFR times} for the estimation of reasonable simulation times, requires the usage of the simulated (and not the experimental) relaxation times.
	
	\subsection{Uncertainty in the dielectric constants through the fluctuation route\label{sec: Uncert FR}}
	In a previous work by the author an others\cite{2018arXiv180306421S}, the uncertainty in the dielectric constant computed through the FR was found. An outline of the derivation of this uncertainty will be done, with some relevant remarks with regard to the individual treatment of each spatial direction. For simulations with typical length, the variance of the sample variance of an autocorrelated random variable, equals to two times the squared variance of the variable divided by the effective degrees of freedom, which can be asymptotically approximated by
	
	\begin{equation}\label{eq: equiv veff}
v_\text{eff} \approxeq  \frac{n}{1+2\sum_{i=1}^{n-1}\phi(i\,\Delta t ) ^2 }
	\end{equation}
	
	The number three in the Eqs. \ref{EQ FR} and \ref{eq: FR estimator} arises by averaging over the three mutually orthogonal spatial directions. The coordinates of the dipole moment are not independent of each others. However, they are not correlated with each other in absence of external fields. Then, applying  the variance operator to Eq. \ref{eq: FR estimator}
	
	\small   
	
	\begin{equation}\label{eq: FR uncert by directions}
	\operatorname{Var}[\varepsilon_r]\approxeq\left(\frac{\operatorname{Var}^2[M_x]}{v_{\text{eff},x}}+\frac{\operatorname{Var}^2[M_y]}{v_{\text{eff},y}}+\frac{\operatorname{Var}^2[M_z]}{v_{\text{eff},z}} \right)\frac{2}{(d\varepsilon_0Vk_BT)^2}
	\end{equation}
	
	\normalsize
	
	\noindent where $d$ represents the number of perpendicular directions considered, and $v_{\text{eff},i}$ the effective degrees of freedom computed for the $i$ direction. Predictably, as the three spatial directions are equivalent to each other, the standard deviation turns to be proportional to $1/\sqrt{d}$ in the long run.

	In analogy to the treatment given for $n_\text{eff}$,
	
	\begin{align}\label{eq: veff and t}
	\nu_\text{eff} &= \lim\limits_{\Delta t \rightarrow 0} \frac{n\Delta t}{\Delta t+2\sum_{i=1}^{n-1}e^{-\frac{2i\Delta t}{\tau}}\Delta t}\\
	&=\frac{\mathcal{T}}{2\int_{0}^{\mathcal{T}}e^{-\frac{2t}{\tau}}d t}=\frac{\mathcal{T}}{\tau}
	\end{align}
	
	\noindent and taking advantage of the three directions, as usual for this calculation route, the following equation is obtained for systems in which the Debye's model is applicable
	
    \begin{equation}\label{eq: FR uncert f(time)}
	\operatorname{Var}[\varepsilon_r]\approxeq    \frac{2\tau}{\mathcal{T}}\frac{\operatorname{Var}^2[\boldsymbol{M}]}{(3\varepsilon_0Vk_BT)^2}
	\end{equation}

	\subsection{Comparison of uncertainties}
In terms of uncertainty, each calculation route has its advantages. They will be compared in terms of the standard deviation of the dielectric constant. As expected, uncertainties are proportional to the inverse of the square root of the total simulation time for both methods (Eqs. \ref{eq: uncert EFR times} and \ref{eq: FR uncert f(time)} ). 

The uncertainty in the dielectric constant obtained through FR does not depends on the system volume. This very significant feature of the FR that has been missed in the literature, as far as the author is aware. The statement above is readily verified by replacing $V$ in Eq. \ref{eq: FR uncert f(time)}, for example, according to Eq. \ref{EQ FR}.

The EFR returns uncertainties inversely proportional to the electric field and the square root of the volume. The  relationship with the field strength and implications are well described in the reference \cite{kolafa2014static}. The increment of the system size comes at expense of computational cost, both magnitudes are often more or less proportional to each other. Although increasing the electric field strength comes for free in terms of computational cost, it should be kept into the limits of validity of Eq. \ref{EQ EFR} when using this framework.

The uncertainties obtained with both methods can be compared by taking their quotient 

\begin{equation}
\frac{\operatorname{Var}[\varepsilon_r]_\text{FR}}{\operatorname{Var}[\varepsilon_r]_\text{EFR}} \approx \frac{\tau}{\tau_E}\frac{E^2\operatorname{Var}^2[\boldsymbol{M}]_{FR}}{(dk_B T)^2\operatorname{Var}[{M}]_{EFR}}\approx \frac{E^2 \varepsilon_0 V (\varepsilon_r-1)}{dk_BT}
\end{equation}

\noindent where it was supposed that $\tau_E\approx\tau$ and that, for each direction, the variance in absence of an external field, e.g. $\operatorname{Var}[{M}_{z}]_{FR}$, equals the variance in the direction of the field of the remaining simulation ($\operatorname{Var}[{M}]_{EFR}$). These approximations are valid only for weak fields. 

The equation above highlights some of the expected qualitative relations between both routes. As $E$ grows, the mean dipole moment increases, and the relative importance of the uncertainty of the mean decreases.
The temperature produces an opposite effect, but this is not significant here due to its change modifies the thermodynamic state for which the dielectric constant is to be computed. Nevertheless, this is significant in choosing a calculation route.

	\subsection{Fluctuation and external field routes combined}
	
	In the case of FR, the three mutually orthogonal spatial directions are used. Instead, only the electric field direction is considered in the EFR. In this work, the possibility of taking advantage of two discarded directions is considered. For them, the probability density function (PDF) remains symmetric. Intuitively, the PDF should be less affected by the field in those directions. Because of this, if the field is weak enough, they can be used for estimating the dielectric constant through the FR, and  the this estimation can be combined with the one corresponding  to the EFR for obtaining a better estimate. 
	
	The plausibility of the intuition above can be illustrated by the results of the following close related problem, which can be solved exactly: the variance of the dipole moment ($\boldsymbol{\mu}=[\mu_x,\mu_y,\mu_z]$) of a rigid dipole in a bath of non-polar particles at constant temperature, which is under the influence of a constant electric field in the $z$ direction. Derivation details were omitted in order to conserve space.  Solving for the $z$ direction,
	
	\begin{align}
	\operatorname{Var}[\mu_z] &= \boldsymbol{\mu}^2\left[\left( \frac{k_BT}{|\boldsymbol{\mu}||\boldsymbol{E}|}\right)^2  - \operatorname{csch}^2\left(\frac{|\boldsymbol{\mu}||\boldsymbol{E}|}{k_BT} \right)   \right]=\\
	&= \frac{\boldsymbol{\mu}^2}{3}-\frac{\boldsymbol{\mu}^2}{15}\left(\frac{|\boldsymbol{\mu}||\boldsymbol{E}|}{k_BT} \right)^2+\dots
	\end{align}

	For $x$ and $y$ directions, 
	
	\begin{align}
	\operatorname{Var}[\mu_x] &=\operatorname{Var}[\mu_y]=\\ 
	 &= \boldsymbol{\mu}^2     \left(\frac{k_BT}{|\boldsymbol{\mu}||\boldsymbol{E}|} \right)^2\left[ \frac{|\boldsymbol{\mu}||\boldsymbol{E}|}{k_BT}\operatorname{coth} \left(\frac{|\boldsymbol{\mu}||\boldsymbol{E}|}{k_BT} \right)-1   \right]=\\
	&= \frac{\boldsymbol{\mu}^2}{3}-\frac{\boldsymbol{\mu}^2}{45}\left(\frac{|\boldsymbol{\mu}||\boldsymbol{E}|}{k_BT} \right)^2+\dots
	\end{align}
	
	It can be seen that for such idealized scenario, the change of the variance in the field direction, due to a weak electric field, is three times greater than the produced in the perpendicular  directions.

	\subsection{An alternative calculation scheme: derivation\label{sec: aleternative approach}}
	
	 Below, a new approach for the calculation of the dielectric constant is presented. It can be briefly described as a three step process: 1. Perform a simulation with no external field. 2. Estimate the expected value that the dipole moment would have if an external field were applied. 3. Employ the equations for the calculation of the dielectric constant of the EFR. Although this method does not provides advantages in its direct application nor significant novelty, it simplifies the study of individual parts when the system is subdivided.

	If a classical system evolves according to a hamiltonian $H_1$, the expected value of an  obervable  $X$ satisfies\cite{tolman1979principles}
	
	\begin{equation}
	\operatorname{E}[\boldsymbol{X}]_{H_1} =\frac{\int _\Omega \boldsymbol{X} e^{-\beta H_1}d\boldsymbol{q}}{\int _\Omega e^{-\beta H_1}d\boldsymbol{q}}
	\end{equation}
	\noindent where $\boldsymbol{q}$ represent the state coordinates, $\Omega$ the domain of the hamiltonian, and  $\beta^{-1}=k_BT$.  The subscript $H_1$ explicit which hamiltonian determines the evolution of the system. If $H_1$ is the hamiltonian of the system under the influence of the external field $\boldsymbol{E}$, and $H_0$ is the hamiltonian of the system with no external field, then $H_1 = H_0 - \boldsymbol{M}\cdot \boldsymbol{E}$\cite{griffiths2005introduction}. Because of this, the expected value of the dipole moment under the influence of the electric field satisfies

    \begin{equation}\label{eq: my meth field}
	\operatorname{E}[\boldsymbol{M}]_{H_1} =\frac{\int _\Omega \boldsymbol{M}  e^{\beta \boldsymbol{M}\cdot \boldsymbol{E}} e^{-\beta H_0}d\boldsymbol{q}}{\int _\Omega e^{\beta \boldsymbol{M}\cdot \boldsymbol{E}} e^{-\beta H_0}d\boldsymbol{q}}=\frac{\operatorname{E}[\boldsymbol{M}  e^{\beta \boldsymbol{M}\cdot \boldsymbol{E}}]_{H_0}}{\operatorname{E}[ e^{\beta \boldsymbol{M}\cdot \boldsymbol{E}}]_{H_0}}
	\end{equation}

	No approximation were made in the derivation of Eq. \ref{eq: my meth field}. However, large perturbations in the MD simulation leads to non representative samples.
	
	\subsection{An alternative calculation scheme: uncertainties}
	
	The uncertainties in the dielectric constant, computed using the procedure described in Section \ref{sec: aleternative approach}, are analyzed bellow. The one-dimensional case is considered in order to facilitate the reading. As before, the uncertainties comes from the fact that the expectation can only be estimated. The problem can be stated as finding a computable  expression for the variance
	
	\begin{align}
	\operatorname{Var}[\varepsilon_r]&=\operatorname{Var}[1+ \frac{\bar{M}_{H_1}}{\varepsilon_r VE}]\\\label{eq: def mean h1}
	\bar{M}_{H_1}&:= \frac{n^{-1}\sum_{i=1}^{n}M_i e^{\beta M_iE}  }{n^{-1}\sum_{i=1}^{n}e^{\beta M_iE}  }
	\end{align}
	
		As noted above, it is required to chose a very weak electric field in order to get a representative sample. In such case, the uncertainty in the computed mean value of the dipole moment will be significant when compared to the obtained estimate of $\operatorname{E}[\boldsymbol{M}]_{H_1}$. A simple solution is to center the values of the sampled dipole moments by subtracting their mean ($\bar M:=\bar M_{H_0}$). This change is small for large values of $n$, as the distribution of $M$ is symmetric for the isotropic systems under study. The Eq. \ref{eq: def mean h1} is modified to  
	
	\begin{equation}\label{eq: my meth centered}
	\bar{M}_{H_1}= \frac{n^{-1}\sum_{i=1}^{n}(M_i - \bar M) e^{\beta (M_i-\bar M)E}  }{n^{-1}\sum_{i=1}^{n}e^{\beta (M_i-\bar M)E}}
	\end{equation}
    
    The denominator of the above equation can be approximated to one by retaining only the first term of its expansion in Maclaurin series. This greatly simplifies the calculation of the variance.

%
%
    \begin{equation}
    n^{-1}\sum_{i=1}^{n}e^{\beta (M_i-\bar M)E} =1+\beta^2 E^2\overline{(M_i-\bar M)^2} +O(M^3) \approx 1
    \end{equation}

The approximation is exact up to first order, the remaining terms can be neglected by using a very weak field.

The numerator can be approximated to a convenient expression

\begin{align}
&n^{-1}\sum_{i=1}^{n}(M_i - \bar M) e^{\beta (M_i-\bar M)E} = \\
&\frac{e^{-\beta \bar M E}}{n}\left[\sum_{i=1}^{n}M_i e^{\beta M_iE}-\sum_{i=1}^{n}\bar M e^{\beta M_iE}   \right]\approx\\
&\frac{1}{n}\left[\sum_{i=1}^{n}M_i e^{\beta M_iE}-\bar M n   \right]=\frac{1}{n}\sum_{i=1}^{n}M_i (e^{\beta M_iE}-1)\label{eq: -1}
\end{align}

Using Taylor expansions for the variance of functions of random variables, the variance of a statistic $f(M)$ can be approximated using\cite{hendebynonlinear}

\small
\begin{equation}
\operatorname{Var}[f(M)] \approx \left( \frac{df(\operatorname{E}[M])}{dM}\right)^2\operatorname{Var}[M]+\frac{1}{2} \left( \frac{d^2f(\operatorname{E}[M])}{dM^2}\right)^2\operatorname{Var}^2[M]
\end{equation}
\normalsize

\noindent then

\begin{equation}
\operatorname{Var}[M_i (e^{\beta M_iE}-1)] \approx (\beta E \operatorname{Var}[M_i])^2
\end{equation}

Here, the estimation of the variance of the mean must take into account that $M$ is auto-correlated, so

\begin{equation}\label{eq: var Mh1}
\operatorname{Var}[\bar{M}_{H_1}] \approx \frac{ (\beta E \operatorname{Var}[M])^2}{n_\text{eff}}
\end{equation}

Finally

\begin{equation}\label{eq my meth uncert}
\operatorname{Var}[\varepsilon_r] \approx \frac{ \beta^2 \operatorname{Var}^2[M]}{(\varepsilon_0V)^2 n_\text{eff}}
\end{equation}

Under the suppositions considered for the equations \ref{eq: neff and t} and \ref{eq: veff and t}, $n_\text{eff}=\nu_\text{eff}/2$. By replacing $n_\text{eff}$, the equation above transform into the one-dimentional case of Eq. \ref{eq: FR uncert by directions}. 

The obtained result is not fortuitous, expanding in Maclaurin series the numerator of Eq. \ref{eq: my meth centered},

\begin{equation}\label{eq: Mh1}
\bar{M}_{H_1} 	\approx   \beta E \overline{(M-\bar M)^2} 
\end{equation}

By inserting this into the Eq. \ref{eq: EFR estimator},  the equation for the calculation of the dielectric constant through the one-dimentional case of FR is obtained.

The absence of the electric field in Eq. \ref{eq my meth uncert} is a consequence of the centering process. If this is not performed, the correct equation is

\begin{equation}
\operatorname{Var}[\varepsilon_r] \approx \frac{ \beta \operatorname{Var}^2[M]}{(\varepsilon_0V)^2 n_\text{eff}} + \frac{ \operatorname{Var}[M]}{(\varepsilon_0VE)^2 n_\text{eff}} 
\end{equation}

\noindent which was derived in the same manner. Predictably, the additional term correspond to the  right hand side of the Eq. \ref{eq: uncert EFR}.

\subsection{Contribution of the individual components}
The knowledge of the individual contributions to dielectric constant of the different parts of a system, may serve to increase the understanding of the interaction among them and the molecular mechanism that give rise to macroscopic properties. Their contribution should be understood in terms of the electric field that they generate, independently of how the other components influenced them. In this sense,  is more natural to express the ideas in terms of the electric susceptibility ($\chi_e = \varepsilon_r-1$), because it can be artificially decomposed in additive contributions, which is not possible for the dielectric constant.

Those contributions can be trivially obtained through the EFR. This is because of the linearity of the expectation operator. For example, if the system is subdivided in electrically neutral parts $A$ and $B$, 

\begin{equation}\label{eq: individual contributions EFR}
\frac{\operatorname{E}[M]}{\varepsilon_0 V E}=\frac{\operatorname{E}[{M}_{A}+{M}_{B}]}{\varepsilon_0 V E} =\frac{\operatorname{E}[{M}_{A}]}{\varepsilon_0 V E}+\frac{\operatorname{E}[{M}_{B}]}{\varepsilon_0 V E}= \varepsilon_r - 1
\end{equation}

\noindent where ${M}_{A}$ and  ${M}_{B}$ represent the component of the dipole moment in the field direction of part $A$ and $B$, respectively.

For the FR this partitioning may not be so obvious. As the variance operator is not linear, in general, $\operatorname{Var}[X_A+X_B] \neq \operatorname{Var}[X_A] + \operatorname{Var}[X_B]$. The contribution of each component, to the variance of the total dipole moment, is obtained by summing over the respective row (or colum) of the covariance matrix. This may be understood in terms of the symmetry of the covariance operator. 
As expected, by summing over every component the expression for the variance of a sum of correlated variables is obtained.

The case of the proposed route of calculation can be treated as follows. Considering the linearity of the expectation operator in Eq. \ref{eq: my meth field},  and that the denominator in Eq. \ref{eq: def mean h1} is approximately one, the mean of the dipole moment of a component $K$ is given by

\begin{equation}
\bar{M}_{K,H_1}\approx \frac{1}{n}\sum_{i=1}^{n} M_{K,i} e^{\beta E \sum_j M_{J,i}}
\end{equation}

The dipole moment of each component in the isotropic system can be centered individually, as the mean of the system is equal to the sum of the mean of its parts. Doing so helps to avoid some bias for individual components. In addition, this leads to the same expression obtained for the FR.

\begin{align}\label{eq: alternative approach individual contribution}
&\bar{M}_{K,H_1}\approx \frac{1}{n}\sum_{i=1}^{n} (M_{K,i}-\bar{M_K}) e^{\beta E \sum_j (M_{J,i}-\bar{M_j})}\\
&\approx\beta E \overline{(M_{K,i}-\bar M_K)^2} +\beta E \sum_{J\neq K} \overline{(M_{K,i}-\bar M_K)(M_{J,i}-\bar M_J)}
\end{align}

Then,  the partial electric susceptibility of the $K$ component is 

\begin{equation}\label{eq: FL uncert contrib}
\frac{\operatorname{Var}[M_{K,i}]+\sum_{J\neq K}\operatorname{Cov}(M_{K,i},M_{J,i})}{\varepsilon_0Vk_BT}
\end{equation}

\noindent This expression coincides with the one expected for the FR. 

\subsection{Uncertainties of the individual contributions}
In this section, the uncertainties in the partial susceptibilities are analyzed. For the EFR, they can obtained  in a manner identical to the employed for the whole system. Then, the variance corresponding to a part of the system, for example, part $A$ is 


\begin{equation}\label{eq: EFR uncert contrib}
\operatorname{Var}\left[\frac{\bar{M}_{A}}{\varepsilon_0 V E} \right]=\frac{  \operatorname{Var}\left[{M}_{A}\right]                }{n_{\text{eff},A}(\varepsilon_0 V E)^2} 
\end{equation}

\noindent where $n_{\text{eff},A}$ is the effective number of observations computed in terms of ${M}_{A}$.


Obtaining an equivalent expression for the FR is a cumbersome process. However, for simulations without an external electric field, the method proposed in Section \ref{sec: aleternative approach} can be used for the present purpose. This is because both methods returns exactly the same values as mentioned in the previous section. The advantage of this approach is that the problem is stated as  obtaining the variance of sample means. If the centering process was performed over every part of the system, Eq. \ref{eq: alternative approach individual contribution} can be expressed as

\begin{equation}
\bar{M}_{K,H_1}\approx \frac{1}{n}\sum_{i=1}^{n} M_{K,i}( e^{\beta E \sum_j M_{J,i}}-1)
\end{equation}

\noindent which was derived using along the same lines that Eq. \ref{eq: -1}.

For the two parts  system, expanding the exponential in Maclaurin series up to second order

\begin{equation}
\bar{M}_{K,H_1}\approx   \frac{\beta E}{n}\sum_{i=1}^{n}   (M_{K,i}^2+ M_{K,i}M_{J,i})
\end{equation}

There is not prejudice in in neglecting the remaining terms of the expansion, as $E$ can be chosen arbitrarily small. Adding up every contribution and applying the expectation operator, the variance of the total dipole moment is obtained up to a constant, see Eq. \ref{eq: Mh1}. In other words, this verifies that the  susceptibility of the whole system computed through the FR can be obtained from the individual contributions if they are individually centered.

The final expression for the uncertainty of the contribution of one part ($K$) of the two parts system, can be obtained using the variable $\bar{M}_{K,H_1}$ in place of $\bar{M}_{A}$ in Eq. \ref{eq: EFR uncert contrib},

\begin{equation}\label{eq: Proposed route uncert contrib}
\frac{  \operatorname{Var}\left[{M}_{K}^2\right]+ \operatorname{Var}\left[{M}_{K}{M}_{J}\right] +2\operatorname{Cov}\left[{M}_{K}^2,{M}_{J}\right] }{n_{\text{eff},K,H_1}   (\varepsilon_0 V kT)^2}
\end{equation}

\noindent where $n_{\text{eff},K,H_1}$ is the effective number of observations computed in terms of the variable $M_{K}^2+ M_{K}M_{J}$. This result can be  generalized for an arbitrary number of parts,

\begin{equation}
\frac{  \operatorname{Var}\left[{M}_{K}^2\right]+ \sum_{J\neq K}\left(\operatorname{Var}\left[{M}_{K}{M}_{J}\right] +2\operatorname{Cov}\left[{M}_{K}^2,{M}_{J}\right]\right) }{n_{\text{eff},K,H_1}   (\varepsilon_0 V kT)^2}
\end{equation}

\noindent where $n_{\text{eff},K,H_1}$ must be evaluated for the variable $M_{K}\sum_{J}M_{J}$.

\section{Numerical experiments}
In order to verify the theoretical development described the previous sections, MD simulations of liquid water were carried out. 

As a first step, the effects of different field strength were considered. Simulations with 512 molecules were performed using the followings values for this property: 0.0005, 0.001, 0.005, 0.01, 0.05, 0.5 and 1 V/nm. The main goal of this step is to test the validity of the equations obtained for the uncertainties, in dielectric constants of the whole system, through the EFR. It is expected for they to be satisfied within the application limit of Eq. \ref{EQ EFR}. A second objective is to verify that the combination of both standard calculation routes may posses practical value. For this to be true, it is required that the FR can be applied for the directions perpendicular to the field, when the latter is large enough to allow the application of the EFR. For the systems studied, a reasonable compromise is found for 0.01 V/nm.

A second step is to analyze the volume dependence of the results. For this purpose, after finding that the electric field strengths of 0.01 V/nm allows the practical application of both routes, simulations with that field strength and the followings numbers of molecules were performed: 256, 1024, 2048, 4096 and 8192.

A third step is to analyze the calculation of the partial susceptibilities and their respective uncertainties. The system studied was again pure water. The advantage over a multicomponent system is that, for the studied system, the contributions of a given part should be more or less proportional to the number of molecules that it contains.

Finally, the relationship between the effective number of samples and the registration time step is briefly analyzed.

\subsection{Simulation details}

The MD simulations were performed in cubic boxes employing the Gromacs 2018.4 \cite{abraham2015gromacs} program. The SPC/E model\cite{berendsen1987missing} was used for water.  The simulation time used is 42 ns, of which the first 2 ns were used for equilibration. 

The time step chosen is 2 fs, unless stated otherwise. Long range coulombic interactions were modeled with PME\cite{darden1993particle}. The temperature was regulated at 298.15 K employing the Bussi-Donadio-Parrinello velocity rescaling algorithm \cite{bussi2007canonical}. The simulation equilibrium pressure was set to 1 bar using the Berendsen's barostat\cite{berendsen1984molecular}. Notice that the applicability of the proposed methods is independent of most simulation parameters.

The posterior analysis was carried out using own Python3 \cite{van2014python} routines based on Numpy\cite{van2011numpy} and MDTraj libraries \cite{mcgibbon2015mdtraj}.

\subsection{Results and discussion}
In order to verify that, in terms of absolute value, the proposed methods provide reasonable uncertainties,  they must be compared with any other procedure. A simple alternative used in this work is explained below.

Consider that the trajectory is evenly split in $p$ contiguous parts of length $\mathcal{T}/p$. According to the equations \ref{eq: uncert EFR times} and \ref{eq: FR uncert f(time)}, the standard deviation in the computed dielectric constant is proportional to $\sqrt{p}$. For each $p$ in $\{p\in\mathbb{N}| 1 < p \leq 100\}$, the dielectric constant of each part can be computed and the standard deviation, of the corresponding $p$ parts, calculated. An estimation of the standard deviation of the dielectric constant, corresponding to the the whole trajectory, is obtained by extrapolating to $p=1$. For the extrapolation, a linear model without the constant term was used. The fitting procedure was performed by means of ordinary least squares.

\begin{figure}[tbh!]
	\centering
	\includegraphics[scale=1]{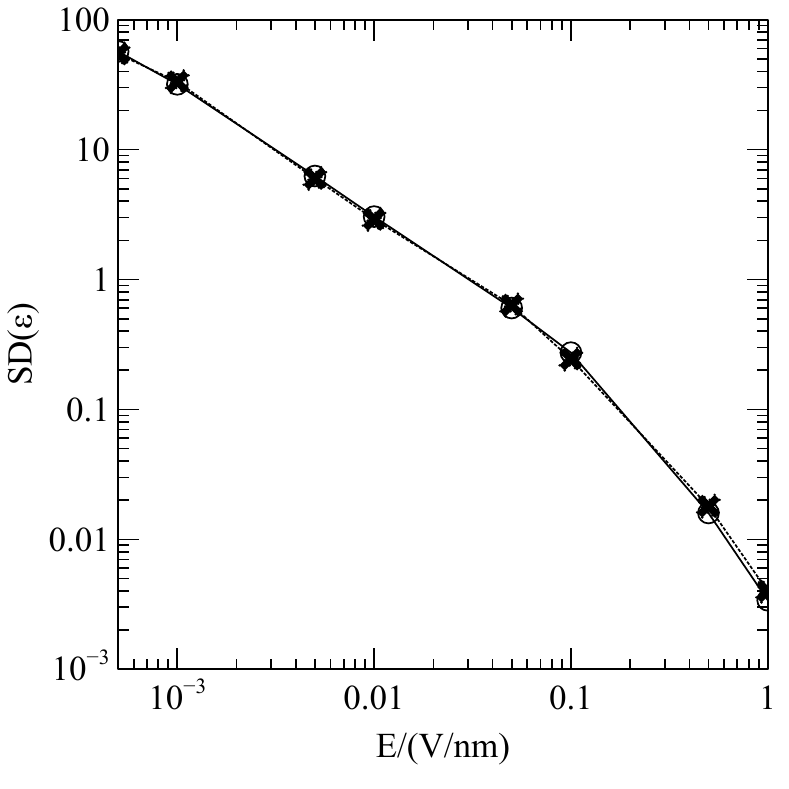}
	\caption{Log-log representation of the estimation of the  deviation in the dielectric constant, computed using the EFR on a system of 512 molecules, vs. the electric field strength in units of V/nm. Empty circles and crosses correspond to the proposed and the numerical method, respectively.}
	\label{fig:fig1}
\end{figure}

The upper part of Table \ref{table1} contains results for the whole system obtained for different values of electric field strength. The estimates corresponding to the EFR can be found in the second column. As expected, for large fields there is a systematic error that decreases the estimate of the dielectric constant value. For these numerical experiments, it starts to be noticeable for electric field of about 0.1 V/nm. The error estimations obtained with the proposed method (Eq. \ref{eq: uncert EFR}) and the numerical method (NM) are found in the third and fourth columns. They are also plotted in Figure \ref{fig:fig1}. This log-log representation present a linear behavior for low fields.

\begin{table*}[tbph!]
	\begin{tabular}{rrrrrrrrr}
		\hline
		$E$/(V/n) &$\varepsilon_{r,\text{EFR}}$& SD$_{\text{EFR,Prop.}}$&  SD$_{\text{EFR,Num.}}$&$\varepsilon_{r,\text{FR},{xy}}$&SD$_{\text{FR}xy}$&SD$_{\text{FR}xy,\text{Num.}    }$&$\varepsilon_{r,\text{mix}}$&SD$_{\text{mix}}$  \\
		0.0005		&55.551	&58.693	&54.455	&69.877	&1.032&1.178	&69.873	&1.031\\
		0.0010		&118.656&31.804	&33.295	&71.238	&1.048&1.007	&71.290	&1.048\\
		0.0050		&57.762	&6.262	&5.996	&71.541	&1.069&1.069	&71.151	&1.053\\
		0.0100		&70.384	&3.051	&2.905	&69.496	&1.033&1.036	&69.587	&0.979\\
		0.0500		&70.471	&0.604	&0.636	&68.891	&1.018&0.997	&70.060	&0.519\\
		0.1000		&67.300	&0.274	&0.244	&69.372	&1.015&0.818	&67.441	&0.264\\
		0.5000		&40.045	&0.016	&0.018	&39.949	&0.435&0.468	&40.045	&0.016\\
		1.0000		&24.794	&0.003	&0.004	&24.797	&0.209&0.183	&24.794	&0.003\\\hline
		Molecules&&&&&&&&|\\
		256 &75.040&4.481&4.359&68.994&1.002&0.960&69.282&0.977\\
		512 &70.384&3.051&2.905&69.496&1.033&1.036&69.587&0.979\\
		1024&68.991&2.134&1.822&72.040&1.075&1.065&71.423&0.960\\
		2048&69.166&1.479&1.441&69.611&1.029&0.915&69.465&0.845\\
		4096&70.589&1.128&1.056&71.672&1.056&1.054&71.166&0.771\\
		8192&69.761&0.778&0.714&70.843&1.050&1.019&70.144&0.625\\
		\hline 
	\end{tabular}
	\centering
	\caption{Results of MD simulations of dielectric constant. Columns 2, 5 and 7 correspond to the EFR, FR in two directions and the MVUE, respectively. Columns 3 and 4 contain the standard deviation (SD) for the EFR through the proposed and the numerical method, respectively. Columns 6, 7 and 9 contain the uncertainties corresponding to the FR and MVUE methods. See main text for details.\label{table1}}
\end{table*}

 In order to compare how well the data obtained is represented by the linear model, the coefficient of determination ($R^2$) of ordinary least square linear regressions of the logarithms of plotted data were calculated. The last two points were removed  because no linear behavior is expected for those strong fields. The  coefficients of determination obtained are 0.9975 and 0.9995 for the proposed and the numerical method, respectively. This mean that the linear model is a good representation of the data.

 The fifth column of the top side of Table \ref{table1}, contains the estimations of the dielectric constant obtained employing the FR for the $x$ and $y$ directions, in which the electric field is null. Their estimated uncertainties are found in the sixth and seventh columns, and were computed using Eq. \ref{eq: FR uncert by directions} and the numerical method, respectively. It can be seen that, in the cases in which the dielectric constant grows linearly with the field, the FR is applicable for the directions in which the electric field is null. This behavior extends to almost 0.1 V/nm. In order to ensure the applicability of the FR, in the following, the value of 0.01 V/nm was used to study the volume dependence of the uncertainties. 

\begin{figure}[tbh!]
 	\centering
 	\includegraphics[scale=1]{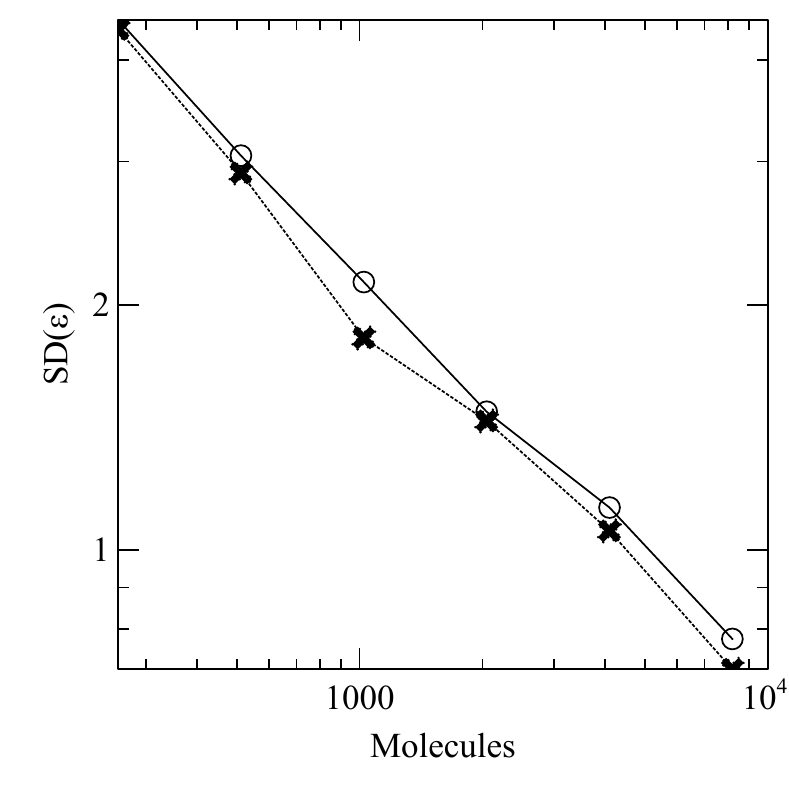}
	\caption{Log-log representation of the estimation of standard deviation in the dielectric constant, computed using the EFR with an electric field of 0.01 V/nm, vs. the number of molecules in the system. Empty circles and crosses correspond to the proposed and the numerical methods, respectively.}
 	\label{fig:fig2}
 \end{figure}

 The lower part of Table \ref{table1} contains the results of simulations performed with the mentioned field, and different numbers of molecules, ranging from 256 to 8192. Figure  \ref{fig:fig2} is a log-log representation of the estimated standard deviations for the dielectric constant computed through the EFR, vs. the number of molecules in the system. For this case, the determination coefficient is 0.9978 for the NM, and 0.9918 for the proposed method (Eq. \ref{eq: uncert EFR}). 

 The standard deviations of the sixth and seventh columns allow to confirm that the volume of the simulation box does not affect the uncertainties in the FR framework. Although well conducted research was done comparing the FR and the EFR\cite{kolafa2014static,riniker2011calculation} or analyzing the behavior of the FR with system size\cite{gereben2011accurate}, this fact has not been made explicit nor suggested. However, the results reported in ref. \cite{gereben2011accurate} clearly support that finding. For the present purposes, it is assumed that the system is large enough for considering that the effects of the periodic boundary conditions are negligible.

 The eighth and ninth columns contain the final estimates of the mean and standard deviation, respectively. They were obtained by employing the minimum-variance unbiased estimator (MVUE)\cite{hartung2011statistical} from the results of both calculation routes. The uncertainties employed are those obtained using the proposed methods. For the sake of completeness, the MVUE was employed in conjunction with the values of all those simulations containing more than 256 molecules, and in which the electric field strength was lower or equal to $0.01$ V/nm. The final result of this estimation is 70.46 $\pm$ 0.31. No systematic error due to system size is expected, as 256 water molecules suffices for reaching the thermodynamic limit in the studied cases\cite{nymand2000molecular}.

In the interests of accuracy, some comments on the effects of the centering process will be made. Both the estimator of Eq. \ref{eq: FR estimator} and the one obtained by removing $\bar{\boldsymbol{M}}^2$ from it have the same expectation for the studied systems. This is because for them $\text{E}[{\boldsymbol{M}}]=\boldsymbol{0}$. However, the estimator of Eq. \ref{eq: FR estimator} is biased. The above was expressed in other terms in Ref.\cite{kolafa2014static}. The estimator of Eq. \ref{eq: FR estimator} has its advantages too; it is invariant under the centering process while the other estimator is not. This suggest that the estimator of Eq. \ref{eq: FR estimator} fits better in the framework of the present study. Despite it is biased, a simple modification allows to build an unbiased estimator\cite{zikeba2010effective}

\begin{equation}
\varepsilon_r = 1 + \frac{n_\text{eff}}{(n_\text{eff}-1)}    \frac{\bar{\boldsymbol{M}^2}-\bar{\boldsymbol{M}}^2}{3\varepsilon_0 Vk_B T}
\end{equation}

The inclusion of the quotient  $n_\text{eff}/(n_\text{eff}-1)$ is not significant in practical terms because, in normal cases, it is very close to one. For example, a value close to 1.0007 is found for the systems simulated in this work. Then, the estimator of Eq. \ref{eq: FR estimator} is perfectly suitable for regular studies. The latter and its unbiased variant were interchangeably used  for the derivation of Eq. \ref{eq: FR uncert by directions}.

The Table \ref{table2} contains the results for the partial susceptibilities, and their respective uncertainties. The system used for the analysis was the one containing 1024 molecules under the influence of an electric field of 0.01 V/nm. For each analyzed case, the system was subdivided into two parts, and one of them was used for the computations. The percentage of the molecules of the system belonging to the analyzed part is tabulated in the first column on the left. When it does not coincide with a natural number of molecules, the last one was truncated. It was verified that the sum of the contributions of both parts coincide with the susceptibility of the whole system. However, it is not true for the uncertainties. The remaining columns contain the calculated contributions and uncertainties. Please, see the caption of Table \ref{table2} for details.

\begin{table*}[tbph!]
	\begin{tabular}{rrrrrrr}
		\hline 
		Percentage &$\chi_{e,\text{EFR}}$& SD$_{\text{EFR,Prop.}}$&  SD$_{\text{EFR,Num.}}$&$\chi_{r,\text{FR},{x}}$&SD$_{\text{FR},x}$&SD$_{\text{FR},x,\text{Num.}}$ \\
		10& 7.189&0.315&0.261& 7.048&0.200&0.188\\
     	25&17.384&0.608&0.507&17.749&0.431&0.402\\
     	50&34.257&1.130&0.941&35.240&0.771&0.703\\
     	75&50.646&1.632&1.382&52.792&1.113&1.030\\
     	90&61.087&1.939&1.662&63.414&1.337&1.231\\     	     	     	
       100&67.991&2.134&1.822&70.401&1.516&1.389\\\hline
	\end{tabular}
	\centering
	\caption{Analysis of the partial susceptibilities. The first column on the left contains the percentage of molecules of the system considered (neglecting rounding errors). The second and third columns contain the estimated partial susceptibilities and their uncertainties computed with the proposed methods for the EFR. The fifth and sixth columns the same for the FR. The remaining columns contain the respective numerical estimations.\label{table2}}
\end{table*}

For a given percentage, many different subsets of molecules could be chosen. In each case, only one subset was used. The results confirm that for both routes of calculation, the fraction of molecules considered approximately equals to the fraction of the susceptibility  of the whole system in which those contribute.

\begin{figure}[tbh!]
	\centering
	\includegraphics[scale=1]{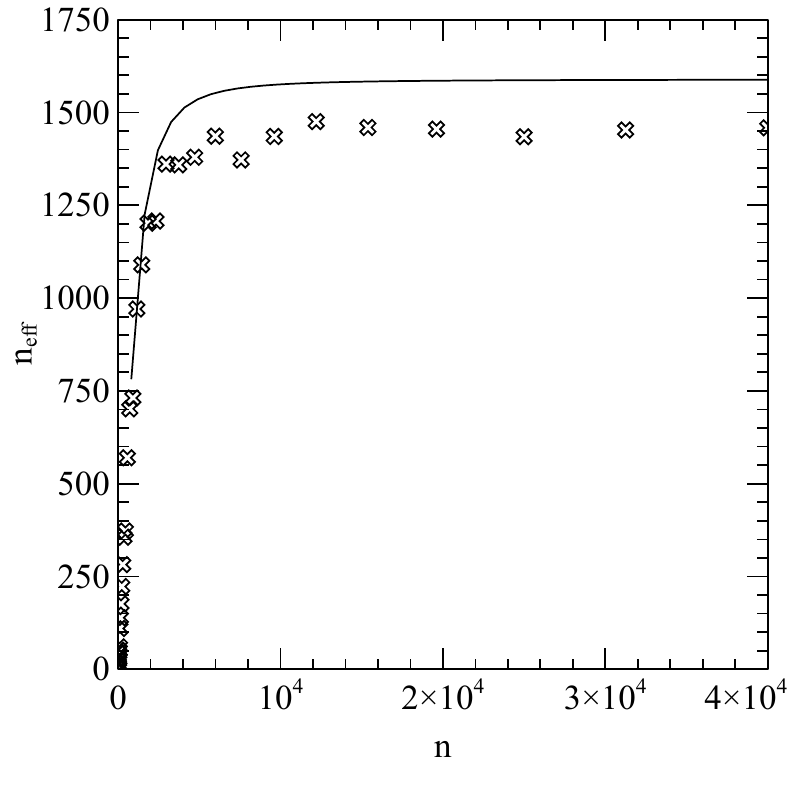}
	\caption{Effective number of samples vs. total number of samples. The continuous line is a plot of  $n\times\operatorname{tanh}[\mathcal{T}(2n\tau)]$ vs. $n$. }
	\label{fig:fig0}
\end{figure}

Another simulation was performed in order to analyze the relationship between the effective number of samples and the registration time step. The simulated system contains 1024 water molecules in presence of an electric field of 0.01 V/m. All the parameters remain the same except that a registration time step of 0.04 ps was used. This implies that the number of samples employed is $10^6$. Only one direction in which the electric field is null was considered. Many pairs $(n, n_{eff})$ were obtained after sub-sampling. Each sub-sample was obtained retaining only those samples whose corresponding ordinals are multiples of some integer. This elemental procedure retrieves samples which may have arisen by using different registration time steps on the same simulation. The pairs obtained are depicted in Figure \ref{fig:fig0}. For comparison, the right hand side of Eq. $\ref{eq: reg freq}$ was plotted.  The required relaxation time (12.6 ps) was computed using the whole sample. Predictably, the effective number of samples increases as the registration time step decreases, and there is not advantage in using  registration time steps much shorter than the relaxation time.  For similar systems, the authors of ref. \cite{gereben2011accurate} found that registration time steps up to around 2 ps are adequate, but a value of 20 ps is not. This is consistent with the analysis and the results presented above. The Eq. \ref{eq: reg freq} can be used for selecting a sampling frequency for the selected system, as can be seen by simple inspection of Figure \ref{fig:fig0}. This is due to the relaxation process of the studied systems is well described by the Debye's model.

\section{Summary}

Two routes for the calculation of the static dielectric constants were studied and compared. One of them is the FR, which consist in sampling the dipole moment of the system in absence of external electric fields, and then using the Eq. \ref{eq: FR estimator}. The expression for the variance of this estimator, previously found by the author and others, was written considering an arbitrary number of directions (Eq. \ref{eq: FR uncert by directions}) and, assuming the applicability of the Debye's model, as a function of the relaxation and simulation times (Eq. \ref{eq: FR uncert f(time)}) . The other route of calculation is the EFR, which requires the application of a constant and external electric field on the system, and employing the estimator of Eq. \ref{eq: EFR estimator}. The variance for this estimator was derived (equations \ref{eq: uncert EFR} and \ref{eq: uncert EFR times}).

The standard deviation of the dielectric constant, obtained through the EFR, is inversely proportional to the electric field. In this work, it was found that the uncertainty does not depends on the volume for the FR, while the dependence is with the inverse of its square root for the EFR.

An alternative calculation scheme was presented. It consist in performing a simulation with no external field, and using Eq. \ref{eq: def mean h1} for the prediction of the mean dipole moment that should have an equivalent simulation, in which the electric field were applied. Then, the result of Eq.  \ref{eq: def mean h1} is used in conjunction with Eq. \ref{eq: EFR estimator} for the calculation of the dielectric constant. When these equations are used after centering the values of the dipole moment, by subtracting its mean, the uncertainty in the dielectric constant is given by Eq. \ref{eq my meth uncert}. The equations above can be applied independently for each spatial direction in order to  lowering the uncertainty. It was also shown that this method reduces to the FR. This fact was used to intuitively show how to compute the partial electric susceptibilities through the FR (Eq. \ref{eq: FL uncert contrib}). Also, their uncertainties were studied for the EFR and the proposed route. It was found that they are represented by Eq. \ref{eq: EFR uncert contrib} and Eq. \ref{eq: Proposed route uncert contrib}, respectively.

The Eq. \ref{eq: equiv neff} is useful for choosing the sampling frequency provided that a small sample is previously available. When the Debye's model is suitable, the Eq. \ref{eq: reg freq} may be used for that purpose. The latter does not requires a previous simulation. One note of caution, however: the relaxation time included in Eq. \ref{eq: reg freq}, which refers to the one obtained through simulation, may differ  significantly from the experimental one.

Te best estimate of the dielectric constant obtained in this work for pure liquid water (SPC/E) at 298.15 K and 1 bar is 70.46~$\pm$~0.31.

\newpage

\bibliography{refs}	

\begin{thebibliography}{10}

\bibitem{von1955dielectric}
Arthur~R von Hippel and SO~Morgan.
\newblock Dielectric materials and applications.
\newblock {\em Journal of The Electrochemical Society}, 102(3):68C--68C, 1955.

\bibitem{asami2002characterization}
Koji Asami.
\newblock Characterization of heterogeneous systems by dielectric spectroscopy.
\newblock {\em Progress in polymer science}, 27(8):1617--1659, 2002.

\bibitem{alder1959studies}
Berni~J Alder and T‌~E Wainwright.
\newblock Studies in molecular dynamics. i. general method.
\newblock {\em The Journal of Chemical Physics}, 31(2):459--466, 1959.

\bibitem{rapaport2004art}
Dennis~C Rapaport and Dennis C~Rapaport Rapaport.
\newblock {\em The art of molecular dynamics simulation}.
\newblock Cambridge university press, 2004.

\bibitem{levrel2008boundary}
L~Levrel and AC~Maggs.
\newblock Boundary conditions in local electrostatics algorithms.
\newblock {\em The Journal of chemical physics}, 128(21):214103, 2008.

\bibitem{de1986computer}
SW~De~Leeuw, John~W Perram, and ER~Smith.
\newblock Computer simulation of the static dielectric constant of systems with
  permanent electric dipoles.
\newblock {\em Annual review of physical chemistry}, 37(1):245--270, 1986.

\bibitem{heinz2001comparison}
Tim~N Heinz, Wilfred~F van Gunsteren, and Philippe~H H{\"u}nenberger.
\newblock Comparison of four methods to compute the dielectric permittivity of
  liquids from molecular dynamics simulations.
\newblock {\em The Journal of chemical physics}, 115(3):1125--1136, 2001.

\bibitem{neumann1983dipole}
Martin Neumann.
\newblock {Dipole moment fluctuation formulas in computer simulations of polar
  systems}.
\newblock {\em Molecular Physics}, 50(4):841--858, 1983.

\bibitem{kolafa2014static}
Jiri Kolafa and Linda Viereblova.
\newblock Static dielectric constant from simulations revisited: Fluctuations
  or external field?
\newblock {\em Journal of chemical theory and computation}, 10(4):1468--1476,
  2014.

\bibitem{riniker2011calculation}
Sereina Riniker, Anna-Pitschna~E Kunz, and Wilfred~F van Gunsteren.
\newblock On the calculation of the dielectric permittivity and relaxation of
  molecular models in the liquid phase.
\newblock {\em Journal of chemical theory and computation}, 7(5):1469--1475,
  2011.

\bibitem{caleman2011force}
Carl Caleman, Paul~J van Maaren, Minyan Hong, Jochen~S Hub, Luciano~T Costa,
  and David van~der Spoel.
\newblock {Force field benchmark of organic liquids: density, enthalpy of
  vaporization, heat capacities, surface tension, isothermal compressibility,
  volumetric expansion coefficient, and dielectric constant}.
\newblock {\em Journal of chemical theory and computation}, 8(1):61--74, 2011.

\bibitem{SANCHEZ2019546}
Hernán~R. Sánchez, Ramiro~M. Irastorza, and C.~Manuel Carlevaro.
\newblock Uncertainties and temperature correction in molecular dynamic
  simulations of dielectric properties of condensed polar systems.
\newblock {\em Journal of Molecular Liquids}, 278:546 -- 552, 2019.

\bibitem{gereben2011accurate}
Orsolya Gereben and L{\'a}szl{\'o} Pusztai.
\newblock On the accurate calculation of the dielectric constant from molecular
  dynamics simulations: The case of spc/e and swm4-dp water.
\newblock {\em Chemical Physics Letters}, 507(1-3):80--83, 2011.

\bibitem{gordon1968correlation}
RG~Gordon.
\newblock {Correlation functions for molecular motion}.
\newblock {\em Advan. Magn. Reson.}, 3:1--42, 1968.

\bibitem{brockwell1991time}
Peter~J Brockwell, Richard~A Davis, and Stephen~E Fienberg.
\newblock {\em Time Series: Theory and Methods: Theory and Methods}.
\newblock Springer Science \& Business Media, 1991.

\bibitem{hilfer2002h}
R~Hilfer.
\newblock {H-function representations for stretched exponential relaxation and
  non-Debye susceptibilities in glassy systems}.
\newblock {\em Physical Review E}, 65(6):061510, 2002.

\bibitem{bottcher1978theory}
Carl Johan~Friedrich B{\"o}ttcher, Oenes~Christoffel van Belle, Paul Bordewijk,
  and Arie Rip.
\newblock {\em Theory of electric polarization}, volume~2.
\newblock Elsevier Science Ltd, 1978.

\bibitem{garrappa2016models}
Roberto Garrappa, Francesco Mainardi, and Maione Guido.
\newblock Models of dielectric relaxation based on completely monotone
  functions.
\newblock {\em Fractional Calculus and Applied Analysis}, 19(5):1105--1160,
  2016.

\bibitem{bayley1946effective}
GV~Bayley and JM~Hammersley.
\newblock {The" effective" number of independent observations in an
  autocorrelated time series}.
\newblock {\em Supplement to the Journal of the Royal Statistical Society},
  8(2):184--197, 1946.

\bibitem{zikeba2010effective}
Andrzej Zieba.
\newblock Effective number of observations and unbiased estimators of variance
  for autocorrelated data-an overview.
\newblock {\em Metrology and Measurement Systems}, 17(1):3--16, 2010.

\bibitem{2018arXiv180306421S}
Hernán~R. Sánchez, Ramiro~M. Irastorza, and C.~Manuel Carlevaro.
\newblock Uncertainties and temperature correction in molecular dynamic
  simulations of dielectric properties of condensed polar systems.
\newblock {\em Journal of Molecular Liquids}, 278:546 -- 552, 2019.

\bibitem{tolman1979principles}
Richard~Chace Tolman.
\newblock {\em The principles of statistical mechanics}.
\newblock Courier Corporation, 1979.

\bibitem{griffiths2005introduction}
David~J Griffiths.
\newblock {\em {Introduction to electrodynamics}}.
\newblock AAPT, 2005.

\bibitem{hendebynonlinear}
Gustaf Hendeby and Fredrik Gustafsson.
\newblock On nonlinear transformations of gaussian distributions.
\newblock 2007.

\bibitem{abraham2015gromacs}
Mark~James Abraham, Teemu Murtola, Roland Schulz, Szil{\'a}rd P{\'a}ll,
  Jeremy~C Smith, Berk Hess, and Erik Lindahl.
\newblock {GROMACS: High performance molecular simulations through multi-level
  parallelism from laptops to supercomputers}.
\newblock {\em SoftwareX}, 1:19--25, 2015.

\bibitem{berendsen1987missing}
HJC Berendsen, JR~Grigera, and TP~Straatsma.
\newblock The missing term in effective pair potentials.
\newblock {\em Journal of Physical Chemistry}, 91(24):6269--6271, 1987.

\bibitem{darden1993particle}
Tom Darden, Darrin York, and Lee Pedersen.
\newblock Particle mesh ewald: An n log (n) method for ewald sums in large
  systems.
\newblock {\em The Journal of chemical physics}, 98(12):10089--10092, 1993.

\bibitem{bussi2007canonical}
Giovanni Bussi, Davide Donadio, and Michele Parrinello.
\newblock {Canonical sampling through velocity rescaling}.
\newblock {\em The Journal of chemical physics}, 126(1):014101, 2007.

\bibitem{berendsen1984molecular}
Herman~JC Berendsen, JPM~van Postma, Wilfred~F van Gunsteren, ARHJ DiNola, and
  JR~Haak.
\newblock Molecular dynamics with coupling to an external bath.
\newblock {\em The Journal of chemical physics}, 81(8):3684--3690, 1984.

\bibitem{van2014python}
Guido {Van Rossum} and Fred~L {Drake Jr}.
\newblock {The Python Language Reference}.
\newblock {\em Python software foundation}, 2014.

\bibitem{van2011numpy}
Stefan Van Der~Walt, S~Chris Colbert, and Gael Varoquaux.
\newblock The numpy array: a structure for efficient numerical computation.
\newblock {\em Computing in Science \& Engineering}, 13(2):22, 2011.

\bibitem{mcgibbon2015mdtraj}
Robert~T McGibbon, Kyle~A Beauchamp, Matthew~P Harrigan, Christoph Klein,
  Jason~M Swails, Carlos~X Hern{\'a}ndez, Christian~R Schwantes, Lee-Ping Wang,
  Thomas~J Lane, and Vijay~S Pande.
\newblock {MDTraj: A modern open library for the analysis of molecular dynamics
  trajectories}.
\newblock {\em Biophysical journal}, 109(8):1528--1532, 2015.

\bibitem{hartung2011statistical}
Joachim Hartung, Guido Knapp, and Bimal~K Sinha.
\newblock {\em Statistical meta-analysis with applications}, volume 738.
\newblock John Wiley \& Sons, 2011.

\bibitem{nymand2000molecular}
Thomas~M Nymand and Per Linse.
\newblock Molecular dynamics simulations of polarizable water at different
  boundary conditions.
\newblock {\em The Journal of Chemical Physics}, 112(14):6386--6395, 2000.

\end{thebibliography}



\end{document}